\documentclass{an}
\usepackage{graphicx}
\usepackage{times}
\usepackage{fancyhdr}
\begin{document}

\title{Revised equipartition \& minimum energy formula for magnetic field
strength estimates from radio synchrotron observations}

\author{Rainer~Beck \and Marita~Krause }

\institute{Max-Planck-Institut f\"ur Radioastronomie,
Auf dem H\"ugel 69, D-53121 Bonn, Germany }

\date{Received 19 April 2005;
accepted 25 May 2005;
published online 1 July 2005}

\abstract{
The commonly used classical equipartition or minimum-energy estimate
of total magnetic fields strengths from radio synchrotron intensities is
of limited practical use because it is based on the hardly known
ratio $\cal K$ of the total energies of cosmic ray protons and electrons
and also has inherent problems. We present a revised formula,
using the \emph{number density ratio} {\bf K} for which we give estimates.
For particle acceleration in strong shocks {\bf K} is about 40
and increases with decreasing shock strength. Our revised estimate for the
field strength gives larger values than the classical estimate
for flat radio spectra with spectral indices of about 0.5--0.6, but smaller values
for steep spectra and total fields stronger than about $10~\mu$G.
In very young supernova remnants, for example, the classical estimate may
be too large by up to 10$\times$. On the other hand, if energy losses of cosmic
ray electrons are important, {\bf K} increases with particle energy
and the equipartition field may be underestimated significantly.
Our revised larger equipartition estimates in galaxy clusters and radio lobes
are consistent with independent estimates from Faraday rotation
measures, while estimates from the ratio between radio synchrotron
and X-ray inverse Compton intensities generally give much weaker fields.
This may be explained e.g. by a concentration of the field in filaments.
Our revised field strengths may also lead to major revisions of electron lifetimes
in jets and radio lobes estimated from the synchrotron break frequency in the
radio spectrum.
}

\correspondence{rbeck@mpifr-bonn.mpg.de}

\headnote{Astron. Nachr./ AN {\bf 326}, No. 6, 414--427 (2005)}

\maketitle
%

\section{Introduction}
\label{intro}

Magnetic fields play an important role in supernova remnants
(F\"urst \& Reich\ \cite{FR04}), in the interstellar medium of
galaxies (Beck et al.\ \cite{B96}; Beck \cite{B04}), in the intergalactic
medium of clusters (Carilli \& Taylor\ \cite{CT02}), and in the jets and
lobes of active galaxies (Rees\ \cite{R85}; Blandford\ \cite{BL01}). The
relative importance among the other competing forces can be estimated by
comparing the corresponding energy densities or pressures.
For most extragalactic objects
measurements of the magnetic field strength are based on radio
synchrotron emission and Faraday rotation measures (RM) of the
polarized radio emission (see Heiles\ \cite{H76} and Verschuur\
\cite{V79} for reviews of the observational methods).

Total synchrotron emission traces the total field in the sky plane,
polarized synchrotron emission the regular field component.
The regular field can be coherent (i.e. preserving its direction within the
region of observation) or incoherent (i.e. frequently reversing its
direction).

Determination of the field strength from the synchrotron intensity needs
information about the number density of the cosmic ray electrons, e.g.
via their $\gamma$-ray bremsstrahlung emission or X-ray emission by
inverse Compton scattering. If such data are lacking, an assumption
about the relation between cosmic ray electrons and magnetic fields
has to be made. The most commonly used approaches are:

\begin{itemize}

\item Minimum total energy density ($\epsilon_\mathrm{tot} =
      \epsilon_\mathrm{CR} + \epsilon_\mathrm{B}= \min$)
\item Equipartition between the total energy densities of cosmic rays
      and that of the magnetic field ($\epsilon_\mathrm{B} = \epsilon_\mathrm{CR}$)
\item Pressure equilibrium ($\epsilon_\mathrm{B} = {1\over3} \epsilon_\mathrm{CR}$)

\end{itemize}
The minimum-energy and equipartition estimates give very similar results and are
often used synonymously. The idea is that cosmic ray particles and magnetic fields
are strongly coupled and exchange energy until equilibrium is reached. Deviations
from equilibrium occur if, for example, the injection of particles or the generation
of magnetic fields is restricted to a few sources or to a limited period.
The mean distance between sources and their mean lifetime define the smallest
scales in space and time over which equipartition can be assumed to hold.

The minimum-energy assumption was first proposed by Burbidge (\cite{B56})
and applied to the optical synchrotron emission of the jet in M87. Since
then the validity of this method has been discussed in the literature.
Duric (\cite{D90}) argued that any major deviation from equipartition
would be in conflict with radio observations of spiral galaxies.

The azimuthally averaged equipartition strength of the field in the Milky Way
and its radial variation (Berkhuijsen, in Beck\ \cite{B01}) agree well with
the model of Strong et al. (\cite{S00}, their Fig.~6) based on radio continuum and
$\gamma$-ray surveys and cosmic ray data near the sun.
On the other hand, Chi \& Wolfendale (\cite{CW93}) claimed significant
deviations from equipartition conditions in the Magellanic Clouds.
Pohl (\cite{P93b}) replied that the standard proton-to-electron
ratio used in the equipartition formula may be smaller in the
Magellanic Clouds compared to the Milky Way.

Equipartition estimates of field strengths were determined in many spiral galaxies
(Beck\ \cite{B00}). The mean strength of the total field ranges from a
few $\mu$G in galaxies with a low star-formation rate to $\simeq30~\mu$G
in grand-design galaxies like M~51. The mean total field strength of a
sample of Shapley-Ames galaxies is $9~\mu$G (Niklas\ \cite{N97}, see
Sect.~\ref{spiral}). The relation between the total field strength and the star
formation rate is deduced from the radio-infrared correlation
(Helou \& Bicay\ \cite{HB93}; Niklas \& Beck\ \cite{NB97}).
The total field is strongest in the spiral
arms, but the strength of the large-scale regular field in most galaxies
is highest in interarm regions (Beck\ \cite{B00}, \cite{B01}).

In the radio lobes of strong FRII radio galaxies the equipartition estimates
of the field strength are 20--100~$\mu$G, in hot spots 100--600~$\mu$G,
and 0.5--1~$\mu$G in the intracluster medium of galaxy clusters and in
relics (see Sect.~\ref{clusters}). In these estimates
relativistic protons were assumed to be absent (``absolute minimum energy'')
or contribute similarly to the total energy density as the electrons.
The ratio of radio synchrotron to inverse Compton X-ray intensities can be
used as another estimate of the field strength (e.g. Carilli \& Taylor\
\cite{CT02}). In most radio lobes the two estimates are similar,
but there are also significant deviations where ``out of equipartition''
conditions have been claimed (see Sect.~\ref{inverse}).

\emph{Faraday rotation measures} RM are sensitive to the coherent regular
field component along the line of sight and to the density of thermal electrons
which can be derived from thermal emission or pulsar dispersion measures DM.
The ratio $RM/DM$ is a widely used measure of coherent field strengths in
the Milky Way. The derived values are lower than the equipartition
estimates from polarized intensity (Beck et al.\ \cite{B+03}).
In galaxy clusters, observations of fluctuations in RM are
used to estimate total field strengths. In the Coma cluster such data
indicate fields one order of magnitude stronger than the equipartition value
(Giovannini et al.\ \cite{G+93}; Feretti et al.\ \cite{F+95}; Fusco-Femiano
et al.\ \cite{F+99}; Govoni \& Feretti\ \cite{GF04}). Strong total fields have
also been claimed from observations of rotation measures in many other clusters
(Carilli \& Taylor\ \cite{CT02}).

All methods to estimate field strengths are subject to bias. Firstly,
a positive correlation between field strength and cosmic ray electron density
on small scales, which are unresolved by the telescope beam or occur
along the line of sight, leads to an \emph{overestimate} of the equipartition
strengths of the total and regular field components (Beck et al.\ \cite{B+03}).
Furthermore, synchrotron intensity is biased towards regions of strong
fields, so that the equipartition estimates are higher than the average
field strength (Sect.~\ref{inverse}).
Field strengths based on Faraday rotation measures may also
be \emph{overestimated} if small-scale fluctuations in magnetic field and
in thermal gas density are positively correlated. Finally, Newman et al.
(\cite{N+02}) and Murgia et al. (\cite{M+04}) pointed out that field
estimates based on Faraday rotation measures are likely to be too high
by a factor of a few if there is a spectrum of turbulence scales.

In this paper we show that there are inherent problems with the classical
equipartition / minimum-energy formula, especially when computing its
parameters from observable quantities.
We present a revised formula which is based on integration of the
cosmic ray energy spectrum rather than the radio frequency spectrum
and discuss the limitations of its application.

\section{The classical minimum-energy formula}
\label{textbook}

The classical textbook minimum-energy formula is based on a value for the
total cosmic ray energy density which is determined by integrating the radio
spectrum from $\nu_\mathrm{\min}$ to $\nu_\mathrm{\max}$,
usually from 10~MHz to 10~GHz, which is the typical range
accessible to observations.
Particles and magnetic fields are assumed to fill the source
homogeneously, and the field is assumed to be completely
tangled (no preferred direction). Then the field minimum-energy
strength $B_{\min}$ is quoted as (e.g. Pacholcyk\ \cite{P70};
Longair\ \cite{L94}, p.~292; Rohlfs \& Wilson\ \cite{RW96}):

\begin{equation}
B_\mathrm{class} = \big(\, 6\, \pi\, G\, ({\cal K}+1)\, L_{\nu}\, / V\, \big)^{2/7}
\label{book}
\end{equation}
where $G$ is a product of several functions varying with
$\nu_\mathrm{\min}$, $\nu_\mathrm{\max}$
and synchrotron spectral index $\alpha$ (see Eq.~(\ref{bmin3}) for details).
$\cal K$ is the ratio of the total energy of cosmic ray
nuclei to that of the synchrotron emitting electrons + positrons. $L_{\nu}$ is the
synchrotron luminosity at frequency $\nu$, $V$ is the source's volume, and
$L_{\nu}/V$ is the synchrotron emissivity.

The resulting magnetic energy density $\epsilon_\mathrm{B}$ is 3/4 of
the cosmic ray energy density $\epsilon_\mathrm{CR}$, so that $B_\mathrm{class}$
is 8\% smaller than the equipartition field strength $B_\mathrm{eq,class}$:

\begin{displaymath}
B_\mathrm{class} / B_\mathrm{eq,class}  =  (3/4)^{2/7}
\end{displaymath}

The problems with the classical formula are the following:
\medskip

(1) The radio emissivity $L_{\nu}/V$ is the average over the source's volume.
Since radio intensity, in the case of equipartition between magnetic fields
and cosmic rays, varies with about the fourth power of the total field
strength, it emerges mainly from the regions with the highest field strengths,
so that $B_{\min}$ is larger than the volume-averaged field strength if the
local field strength $B$ is inhomogeneous within the source.
This problem can be reduced by replacing $L_{\nu}/V$ by $I_{\nu}/l$,
where $I_{\nu}$ is the \emph{local} synchrotron intensity (surface brightness)
and $l$ is the pathlength through the emitting medium (see Eq.~(\ref{bmin3})).
\bigskip

(2) A fixed integration interval $\nu_\mathrm{\min}$ to $\nu_\mathrm{\max}$ is used
in Eq.~(\ref{book}). The critical frequency of synchrotron emission of
an electron with energy $E$ in a magnetic field of strength $B$ with a
component $B_{\perp}$ perpendicular to the velocity vector is
(Lang\ \cite{L99}, p.~29):

\begin{equation}
\nu_\mathrm{crit} \, = \, c_1 \, E^2 \, B_{\perp} \, = \, 16.08~\mbox{MHz} \, (E/\mbox{GeV})^2 \,
(B_{\perp}/\mu\mbox{G})
\label{freq}
\end{equation}
where $c_1=3 e / ( 4 \pi {m_\mathrm{e}}^3 c^5 )$. $e$ is the elementary charge,
$m_\mathrm{e}$ the electron mass, and $c$ the velocity of light.

Note that the synchrotron emission of a single electron is maximum at the
frequency $\nu_\mathrm{max}=0.29\nu_\mathrm{crit}$ (Longair\ \cite{L94}, p.~246), but
for a continuous energy spectrum of electrons the maximum contribution at a given
frequency is emitted by electrons which are a factor of almost two lower in
energy (Webber et al.\ \cite{W80}). As a result, Eq.~(\ref{freq}) can be used
for the \emph{maximum} synchrotron emission from a spectrum of electrons around $E$.

The standard integration interval of 10~MHz--10~GHz corresponds to an interval
[$E_1$ -- $E_2$] in the electron energy spectrum of 800~MeV--25~GeV in a
$1\,\mu$G field, 250~MeV--8~GeV in a $10\,\mu$G field, or to 80~MeV--2.5~GeV in
a $100\,\mu$G field.
The integrated cosmic ray energy $\epsilon_\mathrm{CR}$ is proportional to
[$E_1^{1-2\alpha} - E_2^{1-2\alpha}$] where $\alpha$ is the synchrotron
spectral index (see Eq.~(\ref{energy1})), and $E_1$ and $E_2$ are fixed
integration limits.
Replacing $E_1$ and $E_2$ by $\nu_\mathrm{\min}$ and $\nu_\mathrm{\max}$
in the classical minimum-energy formula
via Eq.~(\ref{freq}) introduces an additional term depending
on the magnetic field strength. As a consequence, the total energy
$\epsilon_\mathrm{tot}$ depends on a constant power of $B$ (see Longair\ \cite{L94},
p.~292) and the wrong derivative $d \epsilon_\mathrm{tot} / d B$
leads to the constant exponent 2/7 in Eq.~(\ref{book}).
The classical minimum-energy formula is {\bf formally incorrect}.
\bigskip

(3) $\cal K$ in Eq.~(\ref{book}) is the ratio of the \emph{total}
energy density of cosmic ray nuclei to that of the electrons (+positrons).
Knowledge of $\cal K$ would require measurements of the spectra of the main
cosmic ray species over the whole energy range, especially at low particle
energies which, for steep spectra, contribute mostly to the total energy.
Instead, the total energy of cosmic ray electrons is approximated in the classical
formula by integrating the radio spectrum between fixed frequency limits,
and the energy of the cosmic ray nuclei is assumed to scale with $\cal K$.

This classical procedure is subject to major uncertainties. Firstly,
the observable radio spectrum only traces the spectrum of cosmic ray electrons
over a small energy range (Eq.~(\ref{freq})). Secondly, the ratio $\cal K$ of
total energies may differ from that in the observable energy range.
What would be needed in the classical formula is the energy ratio $\cal K'$
in the limited energy range traced by the observed synchrotron emission. In case of
energy losses (Sect.~\ref{ratio}) $\cal K'$ may deviate strongly from $\cal K$.
As the other input numbers of the classical formula are generally known with
sufficient accuracy, the uncertainty in $\cal K$ is the reason why the
equipartition / minimum energy field strengths are regarded as crude estimates
and the field estimates are often given with a scaling of $({\cal K}+1)^{2/7}$.
\bigskip

We propose to use, instead of the energy ratio $\cal K$, the
\emph{ratio of number densities} {\bf K} of cosmic ray protons and electrons per
particle energy interval within the energy range traced by the observed
synchrotron emission. Measurements of the local Galactic cosmic rays near the sun
(see Appendix) yield ${\mathrm{\bf K_0}} \simeq100$ at a few GeV, which is the
energy range relevant for radio synchrotron emission. This value is consistent
with the predictions from Fermi shock acceleration and hadronic interaction models
(see Table~\ref{table1}). At lower and higher energies, however, {\bf K} may vary
with particle energy (see Sect.~\ref{ratio}).

The observed energy spectrum of cosmic rays is the result of balance
between injection, propagation and energy losses.
Interactions with matter and radiation are different for protons and electrons,
so that the shape of their energy spectra are generally different
(Pohl\ \cite{P93a}; Lieu et al.\ \cite{LI99}; Schlickeiser\ \cite{S02}).
At low energies (typically below a few 100~MeV) the dominating loss of
cosmic ray protons and electrons is ionization of the neutral gas and/or
Coulomb interaction with the ionized gas.
At energies beyond 1~GeV the dominating loss of protons (and other nucleons) are
inelastic collisions with atoms and molecules, producing pions and secondary
electrons. The spectral index of the stationary energy spectrum is not changed.
The dominating loss mechanism for electrons is nonthermal bremsstrahlung
producing low-energy $\gamma$-rays (Schlickeiser\ \cite{S02}, p.~100).
At even higher energies (beyond a few GeV) the electrons suffer from synchrotron
and inverse Compton losses (see Sect.~\ref{ratio}). Furthermore, the spectra of all
cosmic ray species in galaxies may be steepened if particle diffusion (escape)
is energy-dependent. As the result of energy loss processes, the cosmic ray electron
spectrum is not proportional to the proton spectrum, so that {\bf K}
varies with energy.
Only in a relativistic electron/positron plasma as discussed for jets and
lobes of radio galaxies, where cosmic ray protons are absent, ${\cal K}=0$ and
${\mathrm{\bf K}}=0$ are valid at all energies (see Sects.~\ref{positrons}
and \ref{clusters}).
\bigskip

In this paper we present an easily applicable formula with two input
parameters from observations, synchrotron intensity and spectral index.
We discuss the energy/frequency range where the revised formula can be
applied because a reliable and constant value for the proton-electron number
density ratio ${\mathrm{\bf K_0}}$ can be adopted.
As a result, a more accurate estimate
of the equipartition field strength is possible than in the classical approach.
\bigskip

Pfrommer \& En{\ss}lin (\cite{PE04b}) give two formulae, for the classical
minimum-energy criterion and for the hadronic interaction model applied
to two galaxy clusters, taking into account (2) and (3) discussed above.
Their formula (6.2) for the classical case includes luminosity,
cluster volume, the proton-to-electron energy density ratio and the
lower cut-off energy of the electron spectrum.

\section{The revised formula}

The equipartition and minimum-energy procedures need the total energy of
cosmic rays, to be derived from the observed
radio synchrotron spectrum. An accurate treatment needs to account
for all energy loss processes which affect the energy spectrum of
nucleons and electrons, as discussed in detail by Pohl (\cite{P93a}).
This method, however, requires additional knowledge of the distributions of
gas density and radiation field and is applicable only to a few
well-studied objects.

In this paper we derive a revised formula for the simple case that the number
density ratio ${\mathrm{\bf K_0}}$ of protons and electrons is \emph{constant}
which is valid in a limited range of particle energies. We further
assume that the cosmic rays are accelerated by electromagnetic processes
which generate power laws in momentum, and that the
same total numbers of protons and electrons are accelerated.
The total energy is dominated by the protons. As the proton energy spectrum
flattens below the fixed proton rest energy $E_\mathrm{p}$, the total
cosmic ray energy can be computed easily.
The details are presented in the Appendix.

\subsection{Revised equipartition formula for the total field}
\label{revisedeq}

From the total cosmic ray energy density $\epsilon_\mathrm{CR}$ as a function
of field strength $B$ (Eq.~(\ref{energy7}), see Appendix) and assuming
$\epsilon_\mathrm{CR} = \epsilon_\mathrm{B} = B_\mathrm{eq}^2 /8\pi$ we get:

\begin{eqnarray}
B_\mathrm{eq} & = & \left\{ \, 4\pi (2\alpha+1)\, ({\mathrm{\bf K_0}}+1)\, I_{\nu}\,\,
E_\mathrm{p}^{1-2\alpha}\,\, (\nu/2 c_1)^{\alpha} \right. \nonumber \\
& &\left. \big/ \,\, \big[ (2\alpha-1)\, c_2(\alpha)\, l\, c_4(i) \, \big]
\right\}^{1/(\alpha+3)}
\label{beq}
\end{eqnarray}
$I_{\nu}$ is the synchrotron intensity at frequency $\nu$ and $\alpha$ the
synchrotron spectral index. ${\mathrm{\bf K_0}}$ is the constant ratio of
the number densities
of protons and electrons in the energy range where their spectra are proportional
(see Sect.~\ref{ratio}). $E_\mathrm{p}$ is the proton rest energy.
$c_1$, $c_2$ and $c_4$ are constants. $c_2$ and $c_4$ depend on $\alpha$ and
the magnetic field's inclination, respectively (see the Appendix for
details).

$I_{\nu}$ and $\alpha$ can be determined from observations, while the
proton-to-electron ratio ${\mathrm{\bf K_0}}$ and pathlength $l$
have to be assumed. If the synchrotron sources
have a volume filling factor $f$, $l$ has to be replaced by $l \times f$
in order to obtain $B_\mathrm{eq}$ within the source. $B_\mathrm{eq}$
depends only weakly on the source's distance via the dependence on $l$.

We have restricted the discussion in this paper to nearby sources.
For redshifts $z > 0$, correction terms are required which are given e.g.
in Govoni \& Feretti (\cite{GF04}).

Eq.~(\ref{beq}) yields field strengths which are larger by 7\%
for $\alpha=0.6$, larger by 2\% for $\alpha=0.75$, and smaller by 8\% for
$\alpha=1.0$ compared to the results obtained with the earlier version of the
revised equipartition formula (e.g. Krause et al.\ \cite{K84}; Beck\ \cite{B91};
Niklas\ \cite{N97}; Thierbach et al.\ \cite{T03}). The reason is that a
simplified version of Eq.~(\ref{energy4}) was used previously, with
integration only from $E_1=300$~MeV to $E_2 \rightarrow \infty$.
However, the differences to Eq.~(\ref{beq}) are small, smaller than the
typical errors caused by uncertainties in the input values, so that the
values published by the Bonn group previously are still valid.

In case of equilibrium between magnetic and cosmic ray \emph{pressures},
the field estimate (Eq.~(\ref{beq})) has to be reduced by the factor
$3^{-1/(\alpha+3)}$.

A formula similar to Eq.~(\ref{beq}) has been derived by Brunetti et al.
(\cite{B97}). However, Eq.~(A3) in Brunetti et al., which includes the
lower energy cutoff of the cosmic ray electron spectrum, is not applicable
in the case of dominating protons.

\subsection{Revised minimum energy formula for the total field}
\label{revisedmin}

From Eq.~(\ref{energy7}) (Appendix) and $d\epsilon_\mathrm{tot} / dB \, = \, 0$ we get:

\begin{eqnarray}
B_\mathrm{min} & = & \left\{ \, 2\pi (2\alpha+1)\, (\alpha+1)\, ({\mathrm{\bf K_0}}+1)\, I_{\nu}\,\,
E_\mathrm{p}^{1-2\alpha} \right. \nonumber \\
& & \left. \times (\nu/2 c_1)^{\alpha}\,\, \big/
\big[ (2\alpha-1)\, c_2(\gamma)\, l\, c_4(i)  \big] \right\}^{1/(\alpha+3)}
\label{bmin1}
\end{eqnarray}
The ratio of minimum-energy magnetic and cosmic ray energy densities is:

\begin{displaymath}
\epsilon_\mathrm{B} / \epsilon_\mathrm{CR} = (\alpha+1)/2
\end{displaymath}
Hence, the ratio of minimum-energy (\ref{bmin1}) and equipartition (\ref{beq}) field
strengths is not constant, as in the classical case (\ref{book}), but depends
on the synchrotron spectral index $\alpha$:

\begin{displaymath}
B_\mathrm{min} / B_\mathrm{eq}  = \left[ (\alpha+1)/2\right]^{1/(\alpha+3)}
\end{displaymath}
For $\alpha=1$ the revised formula gives identical results for the
equipartition and minimum energy conditions.

The revised formula is not valid for $\alpha\le0.5$ ($\gamma\le2$)
because the second integral in Eqs.~(\ref{energy1}) and (\ref{energy4}) diverges
for $E_2\rightarrow\infty$. Such flat injection spectra are
observed in a few young supernova remnants and can be explained by
Fermi acceleration operating in a strongly magnetized (low-$\beta$) plasma
(Schlickeiser \& F\"urst\ \cite{SF89}).
In the diffuse medium of galaxies, radio lobes and clusters, $\alpha\le0.5$ is
observed only at low frequencies and indicates strong energy losses of the
electrons due to ionization and/or Coulomb interaction (Pohl\ \cite{P93a};
Sarazin\ \cite{S99}) where the formula cannot be used.

Strong shocks in a non-relativistic $\beta\ge1$ plasma may generate injection
spectra with $\alpha=0.5$ (Schlickeiser \& F\"urst\ \cite{SF89}) where
the total cosmic ray energy is computed according to Eq.~(\ref{energy6}) and the
minimum energy formula has to be modified accordingly:

\begin{eqnarray}
B_\mathrm{min} & = & \left\{4 \pi (\alpha+1) ({\mathrm{\bf K_0}}+1) I_{\nu} E_0^{1-2\alpha}
\left[{1\over2} + \mathrm{ln}(E_2/E_\mathrm{p}) \right] \right. \nonumber \\
& & \left. \times \,\, (\nu/2 c_1)^{\alpha}\,\, \big/\,\,
\big[ c_2(\gamma)\, l\, c_4(i)  \big] \right\}^{1/(\alpha+3)}
\label{bmin2}
\end{eqnarray}

\begin{figure}
\centering
\includegraphics[width=0.45\textwidth]{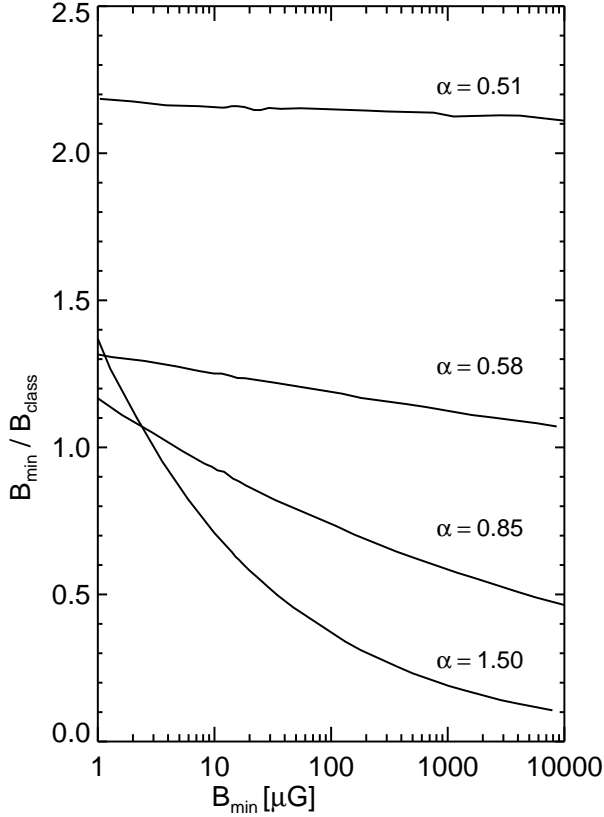}
\caption{Ratio between the revised minimum-energy field strength
$B_\mathrm{min}$ (Eq.~(\ref{bmin1})) and the classical value $B_\mathrm{class}$
(Eq.~(\ref{bmin3}), assuming $\nu_\mathrm{min}=10$~MHz and $\nu_\mathrm{max}=$10~GHz)
as a function of the revised field strength $B_\mathrm{min}$ and
the synchrotron spectral index $\alpha$. A proton-to-electron number
density ratio of ${\mathrm{\bf K_0}}=100$ was adopted.
}
\label{fig1}
\end{figure}

Fig.~\ref{fig1} shows the ratio $q$ between the revised minimum-energy field
strength $B_\mathrm{min}$ (Eq.~(\ref{bmin2})) and the classical value
$B_\mathrm{class}$ (Eq.~(\ref{bmin3})), using ratios of ${\mathrm{\bf K_0}}=100$
and ${\cal K}=100$, respectively. For $\alpha\simeq0.5$ ($\gamma\simeq2$) the ratio
is almost constant and about 2. For larger values of $\alpha$, the ratio
becomes a function of the field strength $B_\mathrm{min}$. For weak fields
(below a few $\mu$G) and values of $\alpha$ between $\simeq0.6$ and $\simeq1$
the revised value differs insignificantly (with respect to the typical
errors of 20\% -- 30\%) from the classical one.

For flat radio spectra ($0.5< \alpha <0.6$) the classical estimate is too low
because the fixed upper limit $\nu_\mathrm{max}=10$~GHz
used for integrating the electron energy excludes the high-energy
part of the cosmic ray spectrum carrying a significant fraction of the total energy.
On the other hand, the classical estimate is too high for steep radio spectra
($\alpha > 0.7$) if the field strength is $> 10~\mu$G. Here the lower integration
limit $\nu_\mathrm{min}=10$~MHz corresponds to energies $<250$~MeV which is
below the lower break energy $E_\mathrm{p}=938$~MeV of the proton spectrum
(see Sect.~\ref{ratio}), so that the total cosmic ray energy is
overestimated.

The ratio $B_\mathrm{min}/B_\mathrm{class}$ depends weakly on
${\mathrm{\bf K_0}}$ ($q\propto {\mathrm{\bf K_0}}^{\,(1/(\alpha+3))\, - 2/7}$)
and on the frequency limits used for the classical formula.
For example, $q$ is about 10\% smaller for $\alpha=0.51$ when
increasing $\nu_\mathrm{max}$ from 10~GHz to 100~GHz.

\subsection{Relativistic electron/positron pair plasma}
\label{positrons}

Jets of radio galaxies may be composed of electrons and positrons
(Bicknell et al.\ \cite{BW01}). Shock acceleration acting on a relativistic
electron/positron pair plasma generates a power law in momentum which leads
to a break in the energy spectrum at
$E_\mathrm{e}=511.00$~keV = $8.1871\cdot 10^{-7}$~erg,
but this break is not observable in the radio range because low-frequency
radio spectra are flattened by ionization and/or Coulomb losses causing
a break at $E_\mathrm{i}$ ($E_\mathrm{i} > E_\mathrm{e}$) in the stationary
energy spectrum (see Sect.~\ref{ratio}).
The revised formulae (\ref{beq}) or (\ref{bmin1}) can be applied to an
electron/positron plasma by using ${\mathrm{\bf K_0}}=0$ and replacing
the lower energy break $E_\mathrm{p}$ by $E_\mathrm{i}$.
If, however, synchrotron or inverse Compton
loss is significant beyond $E_\mathrm{syn}$, the total cosmic ray energy
as to be computed by integration over the observed spectrum, not
according to Eq.~(\ref{energy7}).

The ratio of the revised field strength in a relativistic electron/positron
plasma $B_\mathrm{min,e}$ (${\mathrm{\bf K_0}}=0$) to that in a
proton-dominated plasma $B_\mathrm{min}$ (using Eq.~(\ref{bellratio}) and
assuming $({\mathrm{\bf K_0}}+1)\simeq {\mathrm{\bf K_0}}$) is:

\begin{eqnarray}
B_\mathrm{min,e} / B_\mathrm{min}\,\, & = &
(E_\mathrm{p}/E_\mathrm{e})^{-\alpha_0/(\alpha+3)} \nonumber \\
& & \times \,\, (E_\mathrm{p}/E_\mathrm{i})^{(2\alpha-1)/(\alpha+3)}
\label{bratio}
\end{eqnarray}
where $\alpha_0$ is the spectral index of the injection synchrotron spectrum.
$B_\mathrm{min,e}/B_\mathrm{min}$ varies only weakly with
spectral index $\alpha$ and lower energy break $E_\mathrm{i}$.
For a wide range in $\alpha_0$ and $\alpha$ (0.5--0.75) and in $E_\mathrm{i}$
(10--100~MeV), $B_\mathrm{min,e}/B_\mathrm{min}\simeq 1/3$.

\subsection{Proton-to-electron ratio and energy losses}
\label{ratio}

The ratio {\bf K}{\rm (E)} of the proton-to-electron number densities per
particle energy interval
($n_\mathrm{p}/n_\mathrm{e}$) depends on the acceleration process, the propagation
and the energy losses of protons and electrons. In the range of particle
energies $E_\mathrm{p} < E < E_\mathrm{syn}$, where losses are small or affect
protons and electrons in the same way (``thin target'', see below),
${\mathrm{\bf K}} = {\mathrm{\bf K_0}}$ is constant.

For electromagnetic acceleration mechanisms which generate a power law in
momentum, ${\mathrm{\bf K_0}}$ depends only on the transition energies from
the non-relativistic to the relativistic regime for protons and electrons,
$E_\mathrm{p}$ and $E_\mathrm{e}$, and on the particle injection spectral index
$\gamma_0$ (Eq.~(\ref{number2}); see also Bell\ \cite{B78};
Schlickeiser\ \cite{S02}, p.~472):

\begin{eqnarray}
{\mathrm{\bf K_0}} & = & n_\mathrm{p,0} / n_\mathrm{e,0}\,\,
      = (E_\mathrm{p}/E_\mathrm{e})^{(\gamma_0-1)/2} \nonumber \\
    & = & (E_\mathrm{p}/E_\mathrm{e})^{\alpha_0}\,\,\,\,\,\,\,
      (E_\mathrm{p} < E < E_\mathrm{syn}, \, \mathrm{thin \, target})
\label{bellratio}
\end{eqnarray}
where $\gamma_0$ is the spectral index of the injection cosmic ray spectrum.
$\gamma_0$ is related to the shock compression ratio $r$ via
$\gamma_0=(r+2)/(r-1)$ (see Appendix).
For $\gamma_0\simeq2.2$, as expected from acceleration in supernova remnants,
we get ${\mathrm{\bf K_0}}\simeq100$, consistent with the local Galactic
cosmic ray data near the sun at particle energies of a few GeV (see Appendix).

\begin{figure}
\centering
\includegraphics[bb = 21 16 476 634,angle=270,width=0.45\textwidth]{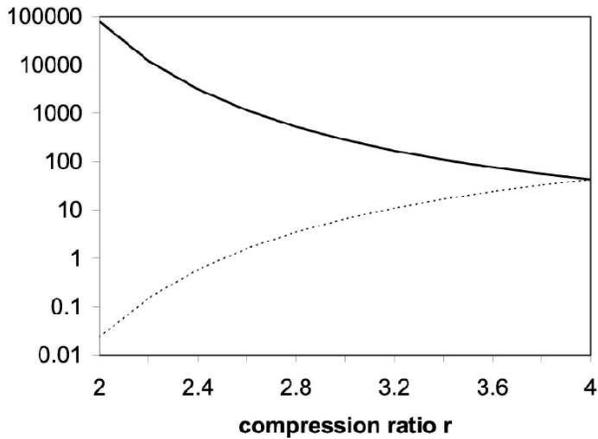}
\caption{Variation of the intrinsic ratio of proton-to-electron number densities
${\mathrm{\bf K_0}}$ (thick line) and of the intrinsic ratio of total energy
densities $\cal K$ (dotted line) with the shock compression ratio $r$.
Energy losses are assumed to be negligible.
}
\label{fig2}
\end{figure}

\begin{center}
\begin{table*}           
\caption{\label{table1} Cosmic-ray proton/electron number density ratio
${\mathrm{\bf K_0}}$
(``thin target'') and the ratio $\cal K$ of total energy densities
for various injection mechanisms}
\begin{tabular}{@{}lll@{}}
\hline
CR origin & ${\mathrm{\bf K_0}}$ & $\cal K$ \\
\hline
\\
Fermi shock acceleration (strong shocks, non-relativistic gas) & 40 -- 100 & 40 -- 20\\
Secondary electrons & 100 -- 300 & $\gg1$\\
Turbulence (resonant scattering of Alfv\'en waves) & $\simeq100$ & $\simeq20$\\
Pair plasma & 0 & 0\\[3pt]
\hline
\end{tabular}
\end{table*}
\end{center}

Eqs.~(\ref{energy2}) and (\ref{bellratio}) are used to compute the intrinsic
ratio $\cal K$ of \emph{total} energies of protons and electrons (for negligible
energy losses):
\begin{equation}
{\cal K} = (E_\mathrm{p}/E_\mathrm{e})^{(3-\gamma_0)/2}
\label{classratio}
\end{equation}
Fig.~\ref{fig2} shows the variation of ${\mathrm{\bf K_0}}$ and $\cal K$ with the
compression ratio $r$ in the shock. Table~\ref{table1} gives values of
${\mathrm{\bf K_0}}$ and $\cal K$ for various injection processes.
For strong shocks in non-relativistic gas ($r=4$, $\gamma_0=2$)
${\cal K}={\mathrm{\bf K_0}}\simeq40$. For steeper intrinsic spectra
(weaker shocks) $\cal K$ decreases, while ${\mathrm{\bf K_0}}$ increases
(Fig.~\ref{fig2}). This demonstrates why the classical formula using $\cal K$
is of little practical use: The number density ratio ${\mathrm{\bf K_0}}$,
in the particle energy range relevant for the observed synchrotron intensity,
is \emph{not} proportional to the energy ratio $\cal K$.

In galaxy clusters secondary electrons are generated by hadronic interactions
between cosmic ray protons and nuclei of the hot gas and may contribute to synchrotron
emission (Dennison\ \cite{D80}). The spectral index $\gamma_\mathrm{e}$ of the
electron spectrum is larger (steeper) than that of the protons $\gamma_\mathrm{p}$,
depending on the model (Pfrommer \& En{\ss}lin\ \cite{PE04a}).
As only a small fraction of the proton energy can be converted into secondary
electrons, the ratio of total proton/electron energies is ${\cal K}\gg 1$.
Assuming balance between injection and
losses, Dennison (\cite{D80}) estimated ${\cal K}=5\,[9 + (B/ \mu G)^2]$.
Typical expected values for the number density ratio in galaxy clusters are
${\mathrm{\bf K_0}}=100$ -- 300 above particle energies of a few GeV
(Pfrommer \& En{\ss}lin\ \cite{PE04b}).
\bigskip

Eq.~(\ref{bellratio}) is valid (i.e. ${\mathrm{\bf K_0}}$ is constant) in the
energy range $E_\mathrm{p} < E < E_\mathrm{syn}$ if the spectral indices
of the electrons and protons are the same. This can be achieved if
energy losses are negligible, or if
the timescale for cosmic ray \emph{escape loss} is smaller than that
for electron loss by nonthermal bremsstrahlung, which is the case for
low interstellar or intergalactic gas densities (``thin target'', Pohl\ \cite{P93a}),
so that nonthermal bremsstrahlung is unimportant for the stationary electron
spectrum. Energy-dependent escape steepens all cosmic ray spectra in
the same way, $\gamma_\mathrm{e}=\gamma_\mathrm{p}=\gamma_0+\Delta\gamma$,
where $-\Delta\gamma\simeq-0.6$ is the exponent of the energy-dependent
escape time $t_{esc}$ (see Appendix), and the revised formula
for equipartition or minimum energy can be applied.

If, however, the gas density is high or the escape time is large, the object
is a ``thick target'' for electrons, so that \emph{nonthermal bremsstrahlung
loss} dominates the electron spectrum at GeV energies. The slopes of the
electron and the radio synchrotron spectra are the same as for the injection
spectra ($\gamma_0$ and $\alpha_0$, respectively). As the loss rate
increases linearly with gas density $n$, the proton-to-electron ratio
increases with gas density.
If the proton spectrum beyond $E_\mathrm{p}$ is steepened by energy-dependent
escape ($t_{esc} \propto E^{-\Delta\gamma}$), e.g. in galaxies, the ratio
{\bf K}{\rm (E)} also depends on energy:

\begin{equation}
{\mathrm{\bf K}(E)} \,\, \propto \, n \, E^{-\Delta\gamma} \,\,\,\,\,\,\,
      (E_\mathrm{p} < E < E_\mathrm{syn}, \, \mathrm{thick \, target})
\label{ratio2}
\end{equation}
The result is that the ratio {\bf K}{\rm (E)} varies over the whole range
of observable energies,
so that \emph{the revised formula for energy equipartition or minimum energy
should not be applied}. The classical formula faces a similar
problem as both the synchrotron intensity and the total energy ratio are
affected by bremsstrahlung loss, but the effect on the ratio $\cal K$ of
total energies is hardly predictable.
\bigskip

We will now discuss the ratio {\bf K}{\rm (E)} for low and high electron
energies (outside the energy range $E_\mathrm{p} < E < E_\mathrm{syn}$).
In a proton-dominated (${\mathrm{\bf K_0}} \gg 1$) cosmic ray plasma,
the lower energy limit
in Eq.~(\ref{bellratio}) or (\ref{ratio2}) is set by the proton spectrum
which flattens below $E_\mathrm{p}=938$~MeV, while the electron spectrum
flattens below $E_\mathrm{e}=511$~keV (Eq.~(\ref{bell})).
For $E < E_\mathrm{p}$ the proton and electron spectra are not proportional,
so that {\bf K}{\rm (E)} is not constant:

\begin{equation}
{\mathrm{\bf K}(E)} = (E/E_\mathrm{e})^{\alpha_0} \,\,\,\,\,\, (E_\mathrm{e} < E < E_\mathrm{p})
\label{kvar1}
\end{equation}
\emph{The revised formula should not be applied} for $E < E_\mathrm{p}$.
In a 10~$\mu$G magnetic field, electrons with $E < E_\mathrm{p}$ are observed at
frequencies below about 140~MHz (Eq.~(\ref{freq})).

The classical formula is also affected if radio observations at low
frequencies are used. Even in a weak 1~$\mu$G magnetic field, the minimum
allowed frequency corresponding to $E_\mathrm{p}$ is still higher than
the standard lower frequency limit of 10~MHz used in the classical formula
(Eq.~(\ref{book})). The standard integration range
of $\ge$10~MHz would trace an appropriate part of the energy spectrum
only if the field strength is below 0.7~$\mu$G. However, synchrotron
emission from such weak fields is too faint to be detected with present-day
radio telescopes. Hence, the lower frequency limit of 10~MHz used
in the classical formula (Eq.~(\ref{book})) is generally too low
and leads to an \emph{overestimate} of the field strength.
This bias increases with increasing field strength (see Fig.~\ref{fig1}).
The ratio $\cal K$ of total energies cannot be applied to
low-frequency radio data.

At low energies, energy losses modify the proton and electron spectra
(Pohl\ \cite{P93a}) and the effective {\bf K}{\rm (E)} is different from
that according to Eq.~(\ref{kvar1}). The break in the electron spectrum by
$\Delta\gamma_\mathrm{e}=-1$ (flattening) due to \emph{ionization and/or
Coulomb loss} depends on gas density (see Eq.~(1) in
Brunetti et al.\ \cite{B97}) and occurs at a few 100~MeV in the ISM of
galaxies and at a few tens MeV in radio lobes and clusters. These energies
correspond to radio frequencies which are generally below the observable
range, so that this effect is irrelevant for equipartition estimates in
a proton-dominated relativistic plasma.
\bigskip

$E_\mathrm{syn}$ in Eq.~(\ref{bellratio}) is the upper energy break where
\emph{synchrotron or inverse Compton loss} of the electrons becomes dominant.
Inverse Compton loss has the same energy and frequency dependence
as synchrotron loss; both are of similar importance if the energy
density of the radiation field and the magnetic energy density are
comparable. Inverse Compton loss from the CMB background dominates if the
field strength is below $3.25\times(1+z)^2~\mu$G.
For $E_\mathrm{e} > E_\mathrm{syn}$ the spectral index steepens by
$\Delta\gamma_\mathrm{e}=1$, observable as a smooth steepening of the
synchrotron spectrum by $\Delta\alpha = 0.5$. In the revised formula for
energy equipartition or minimum energy one should {\bf not} use data at
radio frequencies
corresponding to the electron energy range $E > E_\mathrm{syn}$, where
their energy spectrum is not proportional to the proton spectrum and the
ratio {\bf K} is a function increasing with $E$
(Pohl\ \cite{P93a}; Lieu et al.\ \cite{LI99}). A simplified estimate is:

\begin{equation}
{\mathrm{\bf K}(E)} = {\mathrm{\bf K_0}} (E/E_\mathrm{syn}) \,\,\,\,\,\, (E > E_\mathrm{syn})
\label{kvar2}
\end{equation}
Using instead ${\mathrm{\bf K_0}}$ in the revised formula leads to an
\emph{underestimate} of the field strength.
Again, the classical formula has a similar problem because the
ratio $\cal K$ of total energies also increases in the case of strong synchrotron
or inverse Compton loss.

If the field strength varies along the line of sight or within the volume
observed by the telescope beam, synchrotron loss may lead to an anticorrelation
between field strength and cosmic ray electron density, so that the equipartition
field is underestimated further (Beck et al.\ \cite{B+03}).

Synchrotron loss is significant in sources with strong magnetic fields like
young supernova remnants, starburst galaxies and radio lobes (Sects.~\ref{starburst}
and \ref{clusters}), and also in galaxies away from the acceleration sites of
cosmic rays, for example in interarm regions, in the outer disk
and in the halo (Sect.~\ref{spiral}).

The various bias effects are summarized in Table~\ref{table2}.

\begin{center}
\begin{table*}           
\caption{\label{table2} Bias of equipartition field estimates.
{\bf K}{\rm (E)} is the cosmic ray proton/electron ratio, where
${\mathrm{\bf K_0}}$ denotes a value which does not vary with energy}
\begin{tabular}{@{}lllll@{}}
\hline
Method   &Cosmic-ray composition &Bias effect         &Field strength                 &Reference\\
\hline
\\
Classical &$p^{+}+e^{-}$ (${\mathrm{\bf K_0}}\simeq100$) &Integration over frequency &Underestimate ($\alpha<0.6$) &This paper\\
         &                          &                           &Overestimate ($\alpha>0.7$)  &(Fig.~\ref{fig1})\\[5pt]
Classical &$e^{-} + e^{+}$ (${\mathrm{\bf K_0}}=0$) &Fixed frequency range      &Underestimate (weak fields)  &This paper\\
         &                          &(e.g. 10~MHz--10~GHz)      &Overestimate (strong fields) &           \\[5pt]
Classical+revised&$p^{+}+e^{-}$ ({\bf K}{\rm (E)}$>100$) &Synchrotron/IC/      &Underestimate           &This paper\\
& & bremsstrahlung losses\\[5pt]
Classical+revised& any   &Field fluctuations without/&Overestimate (weak fields)/              &Beck et al.(\cite{B+03})\\
                &       &with synchrotron loss      &Underestimate (strong fields)            &           \\[3pt]
\hline
\end{tabular}
\end{table*}
\end{center}

\subsection{Equipartition and minimum-energy estimates of the regular field}

Knowing the equipartition or minimum-energy estimate of the total field strength,
the equipartition or minimum-energy estimate of the regular field strength
$B_\mathrm{reg,\perp}$ in the sky plane can be
computed from the observed degree of polarization $p$ of the synchrotron emission:

\begin{equation}
p \,\, = \,\, p_0 \, (1 \, + \, \frac{7}{3} q^2) \,\,
/ \,\, (1 \, + \, 3 q^2 \, + \, \frac{10}{9} q^4)
\end{equation}
where $p_0$ is the intrinsic degree of polarization ($p_0=(3-3\alpha)/(5-3\alpha)$)
and $q$ is the ratio of the isotropic turbulent field $B_\mathrm{turb}$
to the regular field component $B_\mathrm{reg,\perp}$ in the sky plane
(Beck et al.\ \cite{B+03}). For the case of a dominating turbulent field ($q\gg1$),
$p\simeq 2.1 \, p_0 \, q^{-2}$.

\section{Discussion and examples}

\subsection{Weak fields: spiral galaxies}
\label{spiral}

The Milky Way and most spiral galaxies have steep radio spectra in the frequency
range of a few GHz which indicates that the energy spectra of their cosmic ray
electrons are steepened by escape loss (see Appendix). Galaxies are
``thin targets'' for cosmic ray electrons, except for massive spiral arms and
starburst regions (see Sect.~\ref{starburst}).
The revised formula can be applied, using the part of the radio spectrum below
the break frequency $\nu_\mathrm{syn}$ beyond which synchrotron or inverse Compton losses
become important. The upper energy limit $E_\mathrm{syn}$ corresponding to
$\nu_\mathrm{syn}$ can be estimated as follows.

The synchrotron lifetime $t_\mathrm{syn}$ for electrons
of energy $E$ is (Lang\ \cite{L99}, p.~32):

\begin{eqnarray}
t_\mathrm{syn} & = & 8.35\cdot 10^9\, \mbox{yr} \left/ \left[ (B_\perp /\mu \mbox{G})^2
\,\, (E_\mathrm{syn}/\mbox{GeV}) \, \right] \right. \nonumber \\
& = & 1.06\cdot 10^9\, \mbox{yr} \left/ \left[ (B_\perp /\mu \mbox{G})^{1.5}
\,\, (\nu_\mathrm{syn}/\mbox{GHz})^{0.5} \,  \right] \right.
\label{tsyn}
\end{eqnarray}
Note that the constant is different from that used in most papers
(e.g. Carilli et al.\ \cite{C+91}; Mack et al.\ \cite{M+98}; Feretti et al.\ \cite{F+98}).
The escape time $t_\mathrm{esc}$ in the Milky Way is $\simeq 10^7$~yr
at non-relativistic and mildly relativistic energies
(Engelmann et al.\ \cite{E+90}; Schlickeiser\ \cite{S02}, p.~439),
so that the break in the radio frequency spectrum is expected at:

\begin{equation}
\nu_\mathrm{syn} \approx 10\, \mbox{GHz} / (B_\perp / 10 \mu\mbox{G})^3
\label{Esyn}
\end{equation}
For the typical strength of ISM magnetic fields of 10~$\mu$G (Niklas\ \cite{N95})
we get $\nu_\mathrm{syn} \simeq 10$~GHz. At higher frequencies the synchrotron
intensity $I_{\nu}$ and the spectral index $\alpha$ can be used neither
for the revised formula nor for the classical formula. If the
resulting field strength is larger than 10~$\mu$G, like in grand-design
spiral galaxies, the useful frequency range shifts to even lower frequencies.
Furthermore, the thermal contribution to the total radio intensity increases
with increasing frequency and has to be subtracted.

In galaxies with a high star-formation rate, galactic winds may form.
The escape time may then be shorter than in the Milky Way and $\nu_\mathrm{syn}$
is larger than according to Eq.~(\ref{Esyn}). Signature of
galactic winds is that at low frequencies the local radio spectrum in
edge-on galaxies remains flat up to large heights above the disk
(Lerche \& Schlickeiser\ \cite{LS82}). In the Milky Way, the radio spectral
index distribution is consistent with a galactic wind (Reich \& Reich\
\cite{RR88}).

Pohl (\cite{P93a}) argued that in the spiral galaxy M~51 the spectral index of the
total radio emission (integrated over the whole galaxy)
is $\alpha\simeq0.5$ ($\gamma_0\simeq2.0$) at low frequencies,
but a frequency break at about 1.4~GHz due to energy losses causes a steepening
of almost the whole observable radio spectrum. As a result, the classical
minimum-energy field strength in the spiral galaxy M~51 is too high.
However, the compilation of all data by Niklas (\cite{N95}) does not
indicate a significant steepening of the radio spectrum of M~51 until
25~GHz, but a flattening due to thermal emission. In his sample of
spiral galaxies, Niklas (\cite{N95}) found only very few cases of spectral
steepening. Hence, escape seems to be the dominant loss process
in the spectra of integrated radio emission of galaxies between about
100~MHz and 10~GHz. (This does not hold for spectra of the \emph{local}
emission, see below.)

Fitt \& Alexander (\cite{FA93}) used the 1.49~GHz radio fluxes of
146 spiral and irregular galaxies to derive minimum-energy field strengths
with the classical formula, simply assuming a constant spectral index of
$\alpha=0.75$, a negligible thermal fraction and the standard frequency limits.
Their distribution peaks around 11~$\mu$G with a standard deviation
$\sigma$ of 4~$\mu$G (for a ratio of total energies $\cal K$ of 100).

Niklas (\cite{N95}) observed a sample of spiral galaxies in radio continuum
at 10.55~GHz and compiled all available data at other frequencies.
Based on the spectra of the integrated radio emission, Niklas et al. (\cite{N97})
could separate the synchrotron from the thermal emission for 74 galaxies.
The mean thermal fraction at 1~GHz is 8\%, and the mean synchrotron
spectral index is $\alpha=0.83$ with a standard deviation of 0.13.
Hence, the average spectrum of the extragalactic cosmic ray electrons
(and probably also that of the protons)
at particle energies of a few GeV has the same spectral index
($\gamma\simeq2.7$) as that in the Milky Way. Niklas (\cite{N95}) derived
galaxy-averaged field strengths according to the classical formula
(\ref{bmin3}) and to a revised formula similar to Eq.~(\ref{bmin1}), approximating
integration (\ref{energy4}) by one integral from $E_1=300$~MeV to infinity.
His results (for ${\mathrm{\bf K_0}}=100$) are similar for the two cases,
as expected from
Fig.~\ref{fig1}. The mean is $9~\mu$G (with a standard deviation
$\sigma=3~\mu$G) both for the revised formula and for the classical case.
The distribution of field strengths is more symmetrical for the
revised case.

However, application of the equipartition formula to regions within spiral
galaxies or in their radio halos needs special care. For example, the
``ring'' of M~31 emitting in radio continuum is formed by cosmic ray
electrons, while the magnetic field extends to regions inside and
outside the ring, as indicated by Faraday rotation data (Han et al.\
\cite{han+98}). Leaving the star-formation regions, cosmic ray electrons
rapidly lose their energy due to synchrotron loss. Hence,
{\bf K}{\rm (E)}$>100$, so that the equipartition formula underestimates
the field strength inside and outside the ring. The same argument holds
for the outer disk and the halo of galaxies, far away from the acceleration
sites of cosmic rays.

In galaxies with strong magnetic fields like M~51, synchrotron loss may
dominate already at a few 100~pc from the acceleration sites, e.g. in
interarm regions, as indicated by spectral steepening between the spiral
arms (Fletcher et al.\ \cite{fletcher+05}). Hence, the interarm field
strengths given by Fletcher et al. (\cite{fletcher+04}) are underestimates.

\subsection{Strong fields: young supernova remnants and starburst galaxies}
\label{starburst}

The classical equipartition formula overestimates the magnetic field strength
in objects with proton-dominated cosmic rays and with strong fields
(Fig.~\ref{fig1}), like massive spiral arms,
young supernova remnants or starburst galaxies.
Here the lower frequency limit is too low and hence overestimates the
total cosmic ray energy (see Sect.~\ref{ratio}).
In a 100~$\mu$G field, for example, the energy of an electron radiating
at 10~MHz is only 80~MeV, where the corresponding proton number in any
medium is strongly reduced ($E \ll E_\mathrm{p}$) and the proton-to-electron
number density ratio {\bf K}{\rm (E)} may drop to values even smaller than 1.

Hence, many values for the field strength quoted in the literature are
strong overestimates. For example, Allen \& Kronberg (\cite{AK98})
integrated the radio spectrum from 10~MHz to 100~GHz of young supernova
remnants in the galaxy M82 and derived field strengths of a few mG for
${\cal K}=0$, scaling with $({\cal K}+1)^{2/7}$. The revised formula gives
values which are between 4 and 50$\times$ smaller, depending on spectral
index $\alpha$. P\'erez-Torres et al. (\cite{P+02}) derived a minimum-energy
field strength in SN~1986J ($\alpha=0.69$) of about
13~mG$\times({\cal K}+1)^{2/7}$ which, according to our revised formula,
is too high by about one order of magnitude.
On the other hand, the equipartition strength of the flat-spectrum SN~1993J
($\alpha=0.51$) of 38~mG (Chandra et al.\ \cite{CRB04}) is too low; the
revised formula gives $\simeq100$~mG$\times({\cal K}+1)^{2/7}$
which fits better to the field strength derived from the synchrotron break
energy (by inserting the SN age as $t_\mathrm{syn}$ in Eq.~(\ref{tsyn})).

Note that the cosmic ray electron distribution in young supernova remnants is not
in a stationary state, so that energy losses modify the spectrum in a
different way than discussed in Sect.~\ref{ratio}. The observed spectrum
is that of injection, possibly modified by synchrotron loss, so that the
revised formula can be applied.

Hunt et al. (\cite{H04}) discussed the radio emission of the blue compact
dwarf galaxy SBS~0335-052. Applying the classical formula (assuming ${\cal K}=40$
and a frequency range of 10~MHz--100~GHz) these authors obtained an
equipartition field of $\simeq0.8$~mG, while the value from our revised
formula is about 30~$\mu$G  which fits much better to other starburst
galaxies like M82 (Klein et al.\ \cite{K88}) or the Antennae
(Chy\.zy \& Beck\ \cite{CB04}) or blue compact
dwarf galaxies (Deeg et al.\ \cite{D+93}).
The same discrepancy may arise in galactic nuclei with starbursts.
Beck et al. (\cite{B+05}) derived field strengths of $\simeq60~\mu$G
in the central starburst regions of the barred galaxies NGC~1097
and NGC~1365, while the classical formula would give much larger values.

However, starburst galaxies and regions of high star formation rate in the
central regions and massive spiral arms of galaxies have high gas densities
and mostly flat radio spectra ($\alpha=0.4$--0.7), so that they probably are
``thick targets'' for the cosmic ray electrons. If so, the equipartition
estimate is too low, and the correct value can be computed only by
constructing a model of gas density and cosmic ray propagation.

\subsection{Radio lobes and galaxy clusters}
\label{clusters}

The classical equipartition field strengths are 20--100~$\mu$G in the radio lobes
of strong FRII radio galaxies (Brunetti et al.\ \cite{B97}), 100--600~$\mu$G in
hot spots (Carilli et al.\ \cite{C+91}; Harris et al.\ \cite{H00}; Wilson et al.\ \cite{W00};
Wilson et al.\ \cite{W01}; Brunetti et al.\ \cite{BB01}; Hardcastle\ \cite{H01};
Hardcastle et al.\ \cite{HB01}; \cite{H02}), and 0.5--1~$\mu$G
in the diffuse gas of galaxy clusters and relics (Feretti \& Giovannini\ \cite{FG96}).
In all these cases it was assumed that relativistic protons are absent
(${\cal K}=0$, ``absolute minimum energy'') or contribute
similarly to the total energy density as the electrons (${\cal K}=1$).

According to Fig.~\ref{fig2}, ${\cal K}\le1$ could also be the result of
a weak shock. However, particle acceleration at low Mach numbers is inefficient
(Bogdan \& V\"olk\ \cite{BV83}; V\"olk et al.\ \cite{V88}), so that its
contribution to the total cosmic ray population is negligible.

If ${\cal K}=0$, the radio spectrum reflects the energy spectrum of the total cosmic
rays, and the equipartition formula
gives reliable results, supposing that the integration limits are set properly
(Sect.~\ref{positrons}).
Integration over the range of Lorentz factors of the electrons/positrons
according to the breaks as observed in the radio spectrum of sources
leads, in the case of weak magnetic fields, to
slightly larger field strengths than applying the classical formula
with fixed frequency limits (Brunetti et al.\ \cite{B97}).
However, the lower break energy $E_\mathrm{i}$ in the electron/positron
energy spectrum needed for the equipartition estimate
may correspond to a radio frequency which is too low to be observed.
For example, electrons with the minimum Lorentz factor of 50 assumed for
the lobes of 3C219 (Brunetti et al.\ \cite{BC99}) in a 10~$\mu$G field
radiate at 100~kHz, a frequency much below the observable range.
Here the frequency limit of $\nu_\mathrm{min}=10$~MHz assumed in the classical
method is too high, and hence the field strength is \emph{underestimated},
by 1.5$\times$ and 2$\times$ for $\alpha=0.8$ and $\alpha=1$, respectively,
or by an even larger factor if the minimum Lorentz factor is smaller than 50.

If the fields are strong, the break energy $E_\mathrm{i}$
of the electron spectrum is observable in the radio spectrum. Hardcastle et al.
(\cite{H02}) determined typical minimum Lorentz factors of 1000 from the
breaks observed at 0.5--1~GHz in hot spots of radio lobes and estimated
field strengths of 100--200~$\mu$G (assuming equipartition with
${\mathrm{\bf K_0}}=0$).
In such cases, the classical formula gives \emph{overestimates} because
$\nu_\mathrm{min}=10$~MHz is too small.

The occurence of electron/positron plasmas (${\cal K}=0$) in astronomical objects
is under debate. A relativistic electron/positron plasma was discussed for
the jets (Bicknell et al.\ \cite{BW01}). However,
observational evidence tends to favour an
electron/proton plasma (Leahy et al.\ \cite{LGT01}). Even if jets do
eject electrons and positrons into a cluster with the same energy spectrum,
these particles will not survive long enough to generate significant
radio emission from the IGM medium, so that (re)acceleration by
intergalactic shocks (Blandford \& Ostriker\ \cite{BO78}; Sarazin\ \cite{S99};
Gabici \& Blasi\ \cite{GB03}),
by interaction with MHD waves (Schlickeiser \& Miller\ \cite{SM98};
Brunetti et al.\ \cite{BSF01}; Fujita et al.\ \cite{FTS03}), or by reconnection
(Hanasz \& Lesch\ \cite{HL03}), or production of secondary electrons
(Pfrommer \& En{\ss}lin\ \cite{PE04a}) is necessary.
\bigskip

If, however, the contribution of relativistic \emph{protons} to the total cosmic
ray energy is dominant (${\cal K}\gg1$), as predicted by all electromagnetic
acceleration models (Table~\ref{table1}),
the equipartition field strength increases with respect to that
for ${\cal K}=0$. In the hot spots of radio lobes
(Wilson et al.\ \cite{W01}; Hardcastle\ \cite{H01}; Hardcastle et al.\ \cite{HB01},
\cite{H02}) stronger equipartition fields would fit much better to the high field
strengths derived from Faraday rotation measures
(Feretti et al.\ \cite{F+95}; Fusco-Femiano et al.\ \cite{F+99}).
\bigskip

Another reason for systematic field underestimates are energy losses.
The electron spectrum is a result of the balance between acceleration and
energy losses. Ionization loss dominates at low energies; at high energies
synchrotron loss dominates in clusters with magnetic fields stronger than
a few $\mu$G while inverse Compton loss due to CMB photons dominates in
clusters with weaker fields (Sarazin\ \cite{S99}).
$E > E_\mathrm{syn}$ is the energy range where synchrotron and/or
inverse Compton losses are strong and the electron spectrum
is steeper than the proton spectrum ($\gamma_\mathrm{e}=\gamma_\mathrm{p}+1$).
The proton-to-electron number density ratio {\bf K} increases with energy
according to Eq.~(\ref{kvar2}). In this case, the revised formula
\emph{underestimates} the field strength. For example,
using a value for {\bf K} of 100 instead of 1000 would underestimate the
equipartition field strength by a factor of about 2.
The classical formula has a similar problem.

Synchrotron and inverse Compton losses steepen the radio spectra.
The radio spectra of most radio lobes and hot spots in the range 0.4--15~GHz
are well approximated by power laws with spectral indices between of
$\alpha\simeq0.7$--0.8 in the frequency range 0.4--15~GHz,
with a spectral break to $\alpha > 1$ between the radio and X-ray regimes.
Meisenheimer et al. (\cite{M89}) fitted the spectra of five bright
hot spots with a low-frequency spectral index of $\alpha\simeq0.5$ with
smooth breaks which start to steepen the spectra already beyond a few GHz.
The spectral index of the diffuse radio emission from the Coma cluster is
$\alpha\simeq 1.0$ below 100~MHz and strongly steepens at higher frequencies
(Fusco-Femiano et al.\ \cite{F+99}; Thierbach et al.\ \cite{T03}),
which is a clear sign of energy loss. The (revised) equipartition field
strength in the Coma cluster is 0.6~$\mu$G$\times({\mathrm{\bf K}}+1)^{1/(\alpha+3)}$
(Thierbach et al.\ \cite{T03}), which gives $\simeq4~\mu$G for $\alpha=0.8$
and ${\mathrm{\bf K}}=1000$. This value is not far from that derived from
Faraday rotation data (Feretti et al.\ \cite{F+95}).

Pfrommer \& En{\ss}lin (\cite{PE04b}) modeled the radio spectrum of the Coma
and Perseus clusters assuming equilibrium between injection of secondary electrons
(hadronic interaction) and energy losses. Their self-consistent minimum-energy
formula, which does not require an assumption about $\cal K$ or
{\bf K}, gives slightly larger
field strengths for the Coma and Perseus clusters compared with the classical
formula. However, this model cannot account for the radial steepening of the radio
spectra observed in both clusters.
\bigskip

Our revision of total field strengths may have significant effects on estimates of
the age of electron populations based on the synchrotron break frequency in the radio
spectrum (Eq.~(\ref{tsyn})). Application of the classical formula for objects with
strong fields may significantly overestimate the field (Fig.~\ref{fig1}) and hence
underestimate the electron lifetime. More serious is the assumption of a
proton/electron total energy ratio ${\cal K}=1$ in most papers (Carilli et al.\ \cite{C+91};
Mack et al.\ \cite{M+98}; Feretti et al.\ \cite{F+98}; Parma et al.\ \cite{P+99})
which is not supported by any acceleration model (see above).
${\mathrm{\bf K_0}}\simeq100$ in the revised formula would
increase field strength by $3.5\times$ and hence decrease the synchrotron age by
$6.5\times$. The discrepancy between dynamical and synchrotron ages found by
Parma et al. (\cite{P+99}) is increased. This problem needs to be re-investigated.

\subsection{Comparison with field estimates from the ratio of synchrotron to
inverse Compton X-ray intensities} \subsectionmark{Comparison with field
estimates ...}
\label{inverse}

The same cosmic ray electrons emitting radio synchrotron emission also produce
X-ray emission via the inverse Compton effect. The ratio of radio to
X-ray intensities can be used as a measure of field strength
(Harris \& Grindlay\ \cite{HG79}; Govoni \& Feretti\ \cite{GF04}).
Comparison with the equipartition estimates reveals three cases:
\bigskip

(1) A high equipartition strength is needed to achieve a consistent
picture.

The lobes of two low-power radio galaxies reveal an apparent deficit
in X-ray emission compared with radio emission when assuming no protons
(${\cal K}=0$)
(Croston et al.\ \cite{C03}). Relativistic protons with 300--500~times more
energy than the electrons would increase the equipartition field strength
from $\simeq3$~$\mu$G to $\simeq15$~$\mu$G and reduce the number density
of electrons, which would explain the weak X-ray emission and also ensure
pressure balance between the radio lobes and the external medium.
\bigskip

(2) A low equipartition strength is needed to achieve a consistent
picture.

The equipartition values (assuming ${\cal K}=0$ or 1)
are similar to those derived from X-ray emission by
inverse Compton scattering of the background photons in most radio lobes
(Brunetti et al.\ \cite{B97}) and in the Coma cluster
(Fusco-Femiano et al.\ \cite{F+99}, see also Table~3 in Govoni \& Feretti\
\cite{GF04}).
In most hot spots the equipartition values are also similar to those
derived from X-ray emission by inverse Compton scattering of the
synchrotron photons (``synchrotron self-Compton emission''), e.g.
in Cyg~A (Harris et al.\ \cite{H94}; Wilson et al.\ \cite{W00}),
3C123 (Hardcastle et al.\ \cite{HB01}), 3C196 (Hardcastle \cite{H01}),
3C295 (Harris et al.\ \cite{H00}), in 3C263 and in 3C330
(Hardcastle et al.\ \cite{H02}).
If, however, ${\cal K}\gg1$, as prediced by acceleration models
(Table~\ref{table1}), the field estimates increase and fit better to the
Faraday rotation data (see above), but contradict the X-ray data in
these objects which then are ``too bright'' in X-rays.
\bigskip

(3) The equipartition strength is always too high.

Some objects are ``too bright'' in X-rays compared with the energy density of the
cosmic ray electrons derived from the equipartition assumption with ${\cal K}=0$.
In other words, the equipartition values even for ${\cal K}=0$ are already
several times larger than those allowed from the ratio of radio/X-ray intensities.
This is the case in the lobes of PKS~1343-601
(Tashiro et al.\ \cite{T98}) and 3C219 (Brunetti et al.\ \cite{BC99}),
and in the hot spots of Pictor~A (Wilson et al.\ \cite{W01}), 3C351
(Brunetti et al.\ \cite{BB01}) and 3C351 (Hardcastle et al.\ \cite{H02}).
Dominating protons (${\cal K}\gg1$) would even increase the
discrepancy.

The correction introduced for strong fields by our revised equipartition formula
(Fig.~\ref{fig1}) cannot solve the problem.
For example, in the bright core of the western hot spot of Pictor A, Wilson et al.
(\cite{W01}) found the largest discrepancy (more than 10) between the field
strength derived from the classical equipartition formula
($\simeq470~\mu$G, assuming ${\cal K}=1$) and that from the radio/X-ray
intensity ratio ($\simeq33~\mu$G). Our revised formula (for ${\mathrm{\bf K_0}}=1$)
reduces the equipartition value only to $\simeq350~\mu$G, not
enough to remove the discrepancy. If cosmic ray protons dominate, the equipartition
field strength \emph{increases} to $\simeq1~$mG (for ${\mathrm{\bf K_0}}=100$
and $\alpha=0.8$), so that the discrepancy between the field estimates
increases further.

Carilli \& Taylor (\cite{CT02}) discussed possible solutions.
Firstly, most of the X-ray emission may not be of inverse Compton, but of
thermal origin, which should be tested with further observations.
Secondly, an anisotropic pitch angle distribution of the cosmic ray
electrons could weaken the synchrotron relative to the inverse Compton
emission. Thirdly, in magnetic fields of about $1~\mu$G the observed
synchrotron spectrum traces electrons of larger energies than those
emitting the inverse Compton spectra. If the electron spectrum
steepens with energy (e.g. due to energy losses), the radio emission is
reduced at high energies.

Another possible (though improbable) explanation of the discrepancy is that
equipartition between magnetic fields and cosmic rays is violated. An increase
in electron density by typically $n'_\mathrm{e}/n_\mathrm{e,eq}\simeq5$ is
required to match the excessive X-ray intensities in the objects listed above.
The corresponding decrease in field strength for a fixed radio synchrotron
intensity and $\alpha\simeq0.8$ is
$B'/B_\mathrm{eq}=(n'_\mathrm{e}/n_\mathrm{e,eq})^{-1/(\alpha+1)}\simeq0.4$, and
the energy density ratio $q$ between cosmic rays and magnetic fields
increases by $q=(n'_\mathrm{e}/n_\mathrm{e,eq})^{(\alpha+3)/(\alpha+1)}\simeq30$.
Such an imbalance between particle and field energies is unstable
and would cause a rapid outflow of cosmic rays.

Finally, the magnetic field may be concentrated in filaments.
Intracluster magnetic fields can be amplified by shocks in merging clusters
(Roettiger et al.\ \cite{RSB99}). The ratio of
synchrotron to X-ray intensities from the same cosmic ray electron spectrum is
$I_\mathrm{syn}/I_\mathrm{X} \, \propto \, <n_\mathrm{e} B^{1+\alpha}> / <n_\mathrm{e}>$.
If small-scale fluctuations in $n_\mathrm{e}$ and in $B$ are uncorrelated,
$I_\mathrm{syn}/I_\mathrm{X} \, \propto \, <B^2> \, = \, <B>^2/f_\mathrm{B}$,
where $f_\mathrm{B}$ is the effective volume filling factor of the magnetic
field, so that the
field strength estimate from the radio/X-ray intensity ratio varies with
$f^{-0.5}$. To explain a 10$\times$ discrepancy between the field estimates,
a filling factor of $10^{-2}$ is required. The enhanced magnetic field in the
filaments leads to synchrotron loss of the electrons and hence to an
anticorrelation between $n_\mathrm{e}$ and $B$. The X-ray emission may be
biased towards regions of weak fields, while synchrotron intensity is biased
towards regions of strong fields. This explanation appears plausible
and deserves further investigation.

\section{Summary}

Assuming that cosmic rays are generated by electromagnetic shock acceleration
producing the same total numbers of relativistic protons and electrons
with power laws in momentum, we showed that the ratio of total energies of protons
and electrons $\cal K$ is about 40 for strong shocks and \emph{decreases} with
decreasing shock strength. The ratio {\bf K} of \emph{number densities}
of protons and electrons per energy interval for particle energies $E\ge1$~GeV is
also about 40 for strong shocks and \emph{increased} with decreasing shock strength.
Both ratios further depend on the various energy loss processes for protons and
electrons.

The classical equipartition or minimum-energy estimate of total magnetic
fields strength from radio synchrotron intensities is based on the ratio $\cal K$
of the total energies of protons and electrons, a number
which is hardly known because the proton and electron spectra have not been measured
over the whole energy range. We propose a revised formula which instead uses the
number density ratio ratio {\bf K} in the energy interval relevant for synchrotron
emission. This ratio can be adopted from the value observed in the local Galactic
cosmic rays,
or it can be estimated for the energy range where energy losses are negligible or
particle escape loss dominates (``thin target''), so that {\bf K} is constant
(${\mathrm{\bf K_0}}$).

Furthermore, the classical equipartition / minimum-energy estimate is
incorrect because the cosmic ray energy density is determined by integration over a
fixed interval in the radio spectrum, which introduces an implicit dependence
on the field strength. Applying our revised formula, the field strengths for a
proton-dominated relativistic plasma are larger by up to 2$\times$ for flat
radio spectra (synchrotron spectral index $\alpha < 0.6$), but smaller by
up to 10$\times$ for steep radio spectra ($\alpha > 0.7$) and for total field
strengths $B>10~\mu$G (Fig.~\ref{fig1} and Table~\ref{table2}).
The revised formula can be applied if energy losses are negligible or if
escape is the dominant loss process for cosmic ray electrons at GeV energies.
The classical field estimates for young supernova remnants
are probably too large by a factor of several.
The average field strengths for spiral galaxies remain almost unchanged.
For objects containing dense gas (``thick targets'') where energy loss of
cosmic ray electrons by nonthermal bremsstrahlung is significant, e.g. in
starburst galaxies, massive spiral arms or in starburst regions in
galaxies, neither the revised nor the classical estimate can be applied.

Equipartition values for radio lobes and galaxy clusters are usually
computed assuming a constant cosmic ray proton/electron energy ratio of ${\cal K}=0$
(i.e. pure electron/positron plasma) or ${\cal K}=1$ (i.e. the same contribution of
protons and electrons to the total cosmic ray energy). However, all current models
of cosmic ray origin predict that protons are dominant (${\cal K}\gg1$,
${\mathrm{\bf K_0}}\simeq100$--300), so that the field estimate is too low by
a factor $({\mathrm{\bf K_0}}+1)^{1/(\alpha+3)}$.
Furthermore, the radio spectra of radio lobes and clusters indicate that
synchrotron or inverse Compton loss of the cosmic ray electrons are significant,
so that $\cal K$ and {\bf K} and hence the field strength increase further.
The revised, stronger fields in clusters are consistent
with the results of Faraday rotation observations.

In case of strong fields in radio lobes, the discrepancy with the much lower field
estimates from the radio/X-ray intensity ratio in several hot spots cannot
be solved and requires alternative explanations, e.g. a concentration of the
field in small filaments with a low volume filling factor, or a thermal
origin of the X-ray emission.

Our code {\sc BFIELD} to compute the strengths of the total and regular fields
from total and polarized synchrotron intensities,
allowing various assumptions as discussed in this paper, is available from
{\em www.mpifr-bonn.mpg.de/staff/mkrause}.

\begin{acknowledgements}
We wish to thank Elly M. Berkhuijsen, Luigina Feretti, Martin Pohl, Wolfgang Reich,
and Anvar Shukurov for many useful discussions, and especially Reinhard Schlickeiser
for his patience to explain to us the labyrinth of cosmic ray loss processes.
Leslie Hunt is acknowledged for encouraging us to publish this paper.
We are grateful to our anonymous referee for his help to make the paper clearer.
\end{acknowledgements}

\begin{appendix}
\section{The revised formula for energy equipartition or minimum energy}
\label{energydensity}

In \emph{all} electromagnetic acceleration models, like the first-order and second-order
Fermi shock acceleration (Drury\ \cite{D83}; Schlickeiser\ \cite{S02}; Gabici \& Blasi\
\cite{GB03}) and the acceleration by MHD waves (Schlickeiser \& Miller\ \cite{SM98};
Brunetti et al.\  \cite{BSF01}; Fujita et al.\ \cite{FTS03}),
the energy spectrum of cosmic rays is a power law in \emph{momentum}
which translates into a power law in \emph{particle energy} $E$ with a break at
$E_\mathrm{b}=m c^2$, where $m$ is the rest mass of the accelerated
particle (Bell\ \cite{B78}):

\begin{eqnarray}
n(E) \, dE & = & n_0 \,\, (E_\mathrm{b}/E_0)^{(1-\gamma_0)/2} \,\,
           (E/E_0)^{-(\gamma_0 +1)/2} \, dE \nonumber \\
           & & (E_1 < E < E_\mathrm{b}) \ , \\
n(E) \, dE & = & n_0 \,\, (E/E_0)^{-\gamma_0} \, dE \nonumber \\
           & & (E > E_\mathrm{b})
\label{bell}
\end{eqnarray}
$n(E)$ is the number of cosmic ray particles per unit volume and per
unit energy interval. $E_0$ ($E_0>E_\mathrm{b}$) is an energy normalization to
obtain correct units; $n_0=n(E_0)$; $E_0$=1~erg if all calculations are performed
in cgs units. $\gamma_0$ is the injection spectral index of the energy spectrum.
$E_1$ is the threshold energy for acceleration (see below).

The solution of the transport equations for Fermi acceleration in a strong,
non-magnetic shock in a non-relativistic gas (compression ratio $r=4$) yields
an injection spectral index of $\gamma_0=(r+2)/(r-1)=2.0$ and $\gamma_0=2.5$
in a relativistic gas (compression ratio $r=3$) (Drury\ \cite{D83};
Schlickeiser\ \cite{S02}). The limit for ultrarelativistic shocks is $\gamma_0=2.23$
(Kirk et al.\ \cite{K00}). In a low-$\beta$ plasma (i.e a dominant magnetic
field, where $\beta$ is the ratio of thermal to magnetic energy densities)
the compression is larger and $\gamma_0$ can be flatter than 2
(Schlickeiser \& F\"urst\ \cite{SF89}). If $\beta$ is close to unity, the
injection spectral index is $\gamma_0\simeq2$, almost independent of the
compression ratio.

In models of first-order Fermi acceleration in supernova shocks the
effective injection spectral index of protons (and probably also electrons)
is $\gamma_0\simeq2.1$--2.3 (Bogdan \& V\"olk\ \cite{BV83}; V\"olk et al.\ \cite{V88})
which agrees well with the source spectrum fitted to the observations
of local Galactic cosmic ray nuclei near the sun (Engelmann et al.\ \cite{E+90}).
Intergalactic shocks due to merging clusters produce electron spectra
with $\gamma_0\simeq2.3$--2.4 (Gabici \& Blasi\ \cite{GB03}).

The total number of cosmic ray particles is (for $\gamma > 1$ and negligible
energy losses):

\begin{eqnarray}
N & = & \int_{E_1}^{\infty} n(E)\, dE \nonumber \\
& = &  \frac{n_0 E_0}{\gamma_0-1} \, [ \, 2 (E_\mathrm{b} E_1 / E_0^2)^{(1-\gamma_0)/2}
  \, - \, (E_\mathrm{b} / E_0)^{(1-\gamma_0)} \, ] \nonumber \\
& \approx & \frac{2 n_0 E_0}{\gamma_0-1} \, (E_0^2 / E_\mathrm{b} E_1)^{(\gamma_0-1)/2}
\end{eqnarray}
\label{number1}
where the approximation is valid for $E_1 \ll E_\mathrm{b}$.

We adopt the commonly used assumption that the same total numbers of
non-relativistic protons and electrons are accelerated above a certain energy
threshold $E_1$, e.g. the non-relativistic kinetic energy of $E_1\simeq 10$~KeV.
From this requirement it follows that the ratio of number
densities at energy $E_0$ ($E_0>E_\mathrm{p}$) has to be:

\begin{equation}
n_\mathrm{p,0} / n_\mathrm{e,0} \, = \, (E_\mathrm{p} / E_\mathrm{e})^{(\gamma_0-1)/2}
\label{number2}
\end{equation}
where $E_\mathrm{p}$ and $E_\mathrm{e}$ are the spectral break energies
for protons and electrons, respectively.\\
($E_\mathrm{p}=938.28$~MeV = $1.5033\cdot 10^{-3}$~erg
and $E_\mathrm{e}=511$~keV = $8.187\cdot 10^{-7}$~erg)\\

The total energy density of cosmic ray particles (for negligible energy losses) is:

\begin{eqnarray}
\epsilon & = & \int_{E_1}^{\infty} n(E)\, E\, dE \nonumber \\
& = & \int_{E_1}^{E_\mathrm{b}} n_0\,
          (E_\mathrm{b}/E_0)^{(1-\gamma_0)/2}\, E_0\,
          (E/E_0)^{(1-\gamma_0)/2}\, dE \nonumber \\
      & & + \int_{E_\mathrm{b}}^{\infty} n_0\, E_0\,
          (E/E_0)^{1-\gamma_0}\, dE
\label{energy1}
\end{eqnarray}
For $E_1 \ll E_\mathrm{b}$ and $2 < \gamma_0 < 3$, the high-energy particles
carry the total energy and Eq.~(\ref{energy1}) gives:

\begin{equation}
\epsilon = n_0 \, E_0^2\, (E_0/E_\mathrm{b})^{\gamma_0-2}
\, ( \frac{2}{3-\gamma_0} + \frac{1}{\gamma_0-2} )
\label{energy2}
\end{equation}
For $\gamma_0=2$ (very strong shocks):

\begin{equation}
\epsilon = n_0\, E_0^2\, ( \, 2 \, + \, \mathrm{ln}(E_2/E_\mathrm{b}) \,)
\label{energy2a}
\end{equation}
where $E_2$ is the high-energy limit of the spectrum according to diffusion
coefficient and/or the finite age of the acceleration region
(Lagage \& Cesarsky\ \cite{LC83}). The TeV $\gamma$-ray image of a
supernova remnant confirmed the existence of particles of very high
energies (Aharonian et al.\ \cite{A04}).
For $\gamma_0=3$:

\begin{equation}
\epsilon = n_0\, E_0^2\, (E_0/E_\mathrm{b})\, ( \, 1 \,
+ \, \mathrm{ln}(E_\mathrm{b}/E_1) \,)
\label{energy2b}
\end{equation}
and for $\gamma_0>3$ (weak shocks) the low-energy particles carry
the total energy:

\begin{eqnarray}
\epsilon & = & n_0\, E_0^2\, (E_0/E_\mathrm{b})^{(\gamma_0-1)/2} \,
 (E_0/E_1)^{(\gamma_0-3)/2} \nonumber \\
& & \times \,\, ( \frac{2}{3-\gamma_0} )
\label{energy2c}
\end{eqnarray}

As a consequence of the different rest masses of protons and electrons,
their spectral break energy $E_\mathrm{b}$ differs, and hence
the energy spectrum of electrons steepens at lower energies than that of
protons (see Fig.~1 in Pohl\ \cite{P93a}).
The proton number per energy interval dominates for $E>E_\mathrm{e}=511$~keV
(see Eqs.~(\ref{bellratio}) and (\ref{kvar1})).

Eq.~(\ref{energy2}) is valid only for the injection spectrum.
The observed spectral index beyond 10~GeV is $\gamma\simeq2.7$ for many cosmic ray
species near the sun (Fulks\ \cite{F75})
and in spiral galaxies (Niklas et al.\ \cite{N97}), which is steeper than the injection
spectrum and indicates an energy-dependent escape time from the Milky Way of
$t_\mathrm{esc}\propto E^{-0.6}$ (Engelmann et al.\ \cite{E+90}). Furthermore, the
sub-relativistic cosmic ray protons suffer from ionization and/or Coulomb
losses, so that the spectrum becomes almost flat at energies below $E_\mathrm{p}$
with an energy spectral index of $(2-\gamma_\mathrm{p,0})/2\simeq0.1$--0.2
(e.g. Pohl\ \cite{P93a}). The observed spectrum of local Galactic protons, corrected for
modulation by the solar wind, gradually steepens between 100~MeV and 10~GeV
(Fulks\ \cite{F75}; Ip \& Axford\ \cite{IA85}) and can be described as:

\begin{eqnarray}
n(E) dE & \cong & \mathrm{const} \,\, dE = n_\mathrm{p,0} \,
(E_\mathrm{p}/E_0)^{-\gamma_\mathrm{p}} \, dE \nonumber \\
           & & (E \la 100\, \mbox{MeV}) \\
n(E) dE & = & n_\mathrm{p,0} \, (E/E_0)^{-\gamma_\mathrm{p}} \, dE \nonumber \\
           & & (E \ga 10\, \mbox{GeV})
\label{energy3}
\end{eqnarray}
with $\gamma_\mathrm{p}\simeq2.7$.
A similar spectral behaviour of the protons is expected in all objects containing
ionized or neutral gas, so that Eq.~(\ref{energy3}) is of general applicability.
Replacing the observed gradual steepening around $E_\mathrm{p}$ by a break at
$E_\mathrm{p}$, the energy density of the protons is approximated by:

\begin{eqnarray}
\epsilon_\mathrm{p} & \cong & \int_{E_1}^{E_\mathrm{p}} n_\mathrm{p,0}\,
         (E_\mathrm{p}/E_0)^{-\gamma_\mathrm{p}}\,  E\,  dE \nonumber \\
& & + \int_{E_\mathrm{p}}^{\infty} n_\mathrm{p,0}\, E_0\, (E/E_0)^{1-\gamma_\mathrm{p}}\, dE
\label{energy4}
\end{eqnarray}
For $E_1 \ll E_\mathrm{p}$ and $\gamma_\mathrm{p} > 2$:

\begin{equation}
\epsilon_\mathrm{p} = n_\mathrm{p,0} \, E_0^2 \, (E_0/E_\mathrm{p})^{\gamma_\mathrm{p}-2}\,\,
({1\over2} + \frac{1}{\gamma_\mathrm{p}-2} )
\label{energy5}
\end{equation}

For very strong shocks which generate injection spectra with $\gamma_\mathrm{p}=2.0$
(Sect.~\ref{revisedmin}), Eq.~(\ref{energy5}) has to be replaced by:

\begin{equation}
\epsilon_\mathrm{p} = n_\mathrm{p} E_0^2\,
({1\over2} + \, \mathrm{ln}(E_2/E_\mathrm{p}) \,)
\label{energy6}
\end{equation}

The scaling parameter $n_\mathrm{p,0}$ of the proton spectrum can be determined
in ``thin targets'' from that of the electron spectrum $n_\mathrm{e,0}$ and the
proton-to-electron number density ratio ${\mathrm{\bf K_0}}$, and $n_\mathrm{e,0}$
can be replaced by the synchrotron intensity
(surface brightness) $I_{\nu}$ (in $\rm erg\,\, s^{-1}\, cm^{-2}\, Hz^{-1}\,
sterad^{-1}$), observed at the frequency $\nu$, and the magnetic field strength
$B_{\perp}$ in the sky plane (in G), making use of the synchrotron formula
(e.g. Pacholczyk\ \cite{P70}):

\begin{equation}
I_{\nu} = c_2(\gamma_\mathrm{e})\,\, n_\mathrm{e,0}\,\, E_0^{\gamma_\mathrm{e}}\,\, l\,\,
(\nu/2c_1)^{(1-\gamma_\mathrm{e})/2}\,\,\, B_{\perp}^{(\gamma_\mathrm{e}+1)/2}
\label{int}
\end{equation}
where
\begin{eqnarray}
\nonumber
c_1         & = & 3e/(4\pi {m_\mathrm{e}}^3 c^5) = 6.26428\cdot 10^{18}\,\,
                 \mbox{erg$^{-2}$ s$^{-1}$ G$^{-1}$}, \nonumber \\
c_2(\gamma_\mathrm{e}) & = & {1\over4} c_3\,(\gamma_\mathrm{e}+7/3)/
(\gamma_\mathrm{e}+1) \, \Gamma [(3\gamma_\mathrm{e}-1)/12] \nonumber \\
& & \times \, \Gamma [(3\gamma_\mathrm{e}+7)/12] \nonumber \\
c_3         & = & \sqrt{3}\, e^3 / (4\pi m_\mathrm{e} c^2) \nonumber \\
& = & 1.86558\cdot 10^{-23}\,\, \mbox{erg\, G$^{-1}$ sterad$^{-1}$} \nonumber
\end{eqnarray}
$c_2(\gamma_\mathrm{e})$ is identical to $c_5(\gamma_\mathrm{e})$ in Pacholczyk
(\cite{P70}, p.~95 and tabulated on p.~232). $n_\mathrm{e}(E)$
is the number density of cosmic ray electrons per unit energy interval
(in cm$^{-3}$ erg$^{-1}$) in the relevant energy range.
(Note that Pacholczyk (\cite{P70}) used a different definition of
$n_\mathrm{e}(E)$ which leads to strange units.)
$l$ is the pathlength through the emitting medium.
The spectral index $\alpha$ of the synchrotron
emission relates to the spectral index $\gamma_\mathrm{e}$ of the energy spectrum
of the electrons via $\alpha=(\gamma_\mathrm{e}-1)/2$.

We define $E_\mathrm{syn}$ as the electron energy beyond which synchrotron
and inverse Compton losses are significant, hence the ratio {\bf K}{\rm (E)}
is not constant, and the
equipartition estimate fails (see Sect.~\ref{ratio}). In the energy range
$E_\mathrm{p} < E < E_\mathrm{syn}$, the ratio is constant
(${\mathrm{\bf K}} = {\mathrm{\bf K_0}}$) and the proton and
electron spectra are power laws with the same spectral index
(${\gamma_\mathrm{e}}={\gamma_\mathrm{p}}={\gamma}$) if
nonthermal bremsstrahlung loss of the electrons is negligible (see Sect.~\ref{ratio}).
Then the radio synchrotron spectrum can be used to extrapolate the proton spectrum,
so that $n_\mathrm{p}$ in Eq.~(\ref{energy5}) can be replaced by:

\begin{displaymath}
n_\mathrm{p,0} = {\mathrm{\bf K_0}} \, I_{\nu}\, (\nu/2c_1)^{(\gamma-1)/2} \left/
     \, [ \, c_2(\gamma)\, E_0^{\gamma}\, l\, B_\perp^{(\gamma+1)/2} \, ] \right.
\end{displaymath}
and the total cosmic ray energy density becomes:

\begin{eqnarray}
\epsilon_\mathrm{CR} & = & \gamma \, ({\mathrm{\bf K_0}}+1)\, I_{\nu}\,\, E_\mathrm{p}^{2-\gamma}\,\,
     (\nu/2c_1)^{(\gamma-1)/2} \nonumber \\
& & \big/ \, [ \, 2 (\gamma-2)\, c_2(\gamma)\, l\,
     B_\perp^{(\gamma+1)/2} \, ]
\label{energy7}
\end{eqnarray}
Note that, according to Eq.~(\ref{energy2}), $\epsilon_\mathrm{CR}$ includes protons
up to the highest energies.

The projected field component $B_\perp$ has to be replaced by the total
field $B$:

\begin{equation}
B_\perp^{(\gamma+1)/2} = B^{(\gamma+1)/2} c_4(i)
\label{geo1}
\end{equation}
$c_4(i) = [cos(i)]^{(\gamma+1)/2}$ is valid for observation of a region
where the field is completely regular and has a constant inclination $i$ with
respect to the sky plane ($i=0^o$ is the face-on view), e.g. in a small region of
the Milky Way or a galaxy. If the field is completely turbulent and has an
isotropic angle distribution in three dimensions, $B_\perp^2 = (2/3) B^2$
and $c_4(i) = (2/3)^{(\gamma+1)/4}$.
If the synchrotron intensity is averaged over a large volume,
$[cos(i)]^{(\gamma+1)/2}$ has to be replaced by its average over
all occuring values of $i$. The case of a field oriented parallel to a
galactic disk was treated by Segalovitz et al. (\cite{S76}, their Eq.~(3)).

Combining $\epsilon_\mathrm{CR} = \epsilon_\mathrm{B} = B_\mathrm{eq}^2 /8\pi$
with Eq.~(\ref{energy7}) and replacing $\gamma$ by ($2\alpha+1$) leads to:

\begin{eqnarray}
B_\mathrm{eq} & = & \left\{ \, 4\pi (2\alpha+1)\, ({\mathrm{\bf K_0}}+1)\, I_{\nu}\,\,
E_\mathrm{p}^{1-2\alpha}\,\, (\nu/2 c_1)^{\alpha} \right. \nonumber \\
& & \left. \big/\,\, \big[ (2\alpha-1)\, c_2(\alpha)\, l\, c_4(i) \, \big]
\right\}^{1/(\alpha+3)}
\label{beqA}
\end{eqnarray}

For use in the classical formula, Eq.~(\ref{geo1}) has to be replaced by:

\begin{equation}
B_\perp^{3/2} = B^{3/2} c_4(i)
\label{geo2}
\end{equation}

With the definitions introduced before,
we can write the classical minimum-energy formula (\ref{book}) as:

\begin{eqnarray}
B_\mathrm{class} & = & \left\{ \, 6\pi ({\cal K}+1)\, I_{\nu}\,\,
(\nu/2)^{\alpha}\,\, c_1^{-1/2}\,\, \big[ \nu_\mathrm{min}^{(1/2\, -\alpha)} \right. \nonumber \\
& & \left. - \nu_\mathrm{max}^{(1/2\, -\alpha)} \big] \,\, \big/ \,\,
\big[ (2\alpha-1)\, c_2(\alpha)\, l\, c_4(i)  \big] \right\}^{2/7} \
\label{bmin3}
\end{eqnarray}
A table to compute $B_\mathrm{class}$ from $I_{\nu}$ (in mJy/arcsec$^2$)
and $l$ (in kpc) is given in Govoni \& Feretti (\cite{GF04}).

As discussed in Sect.~\ref{textbook}, the integration of the radio spectrum
between the fixed frequencies $\nu_\mathrm{min}$ and $\nu_\mathrm{max}$
introduces an implicit dependence on field strength $B$ which is ignored when
computing the derivative $d \epsilon_\mathrm{tot} / d B$.
The constant exponent 2/7 in Eq.~(\ref{bmin3}) is the result of this error,
which is one of the reasons why the classical formula gives wrong results
(see Sect.~\ref{textbook}).

\end{appendix}
\end{document}